\documentclass[twocolumn,aps,prb,superscriptaddress]{revtex4}
\usepackage{graphicx}
\usepackage{color}
\usepackage{amsmath}
\usepackage{enumitem}
\usepackage{amssymb}
\usepackage{hyperref}
\usepackage[normalem]{ulem}

\hypersetup{
	colorlinks = true,
	linkcolor = blue,
	citecolor = blue
}

\newcommand{\be}{\begin{equation}}
\newcommand{\ee}{\end{equation}}

\newcommand{\bea}{\begin{eqnarray}}
\newcommand{\eea}{\end{eqnarray}}

\newcommand{\lp}{\left(}
\newcommand{\rp}{\right)}

\renewcommand{\vec}[1]{{\boldsymbol #1}}
\renewcommand{\epsilon}{\varepsilon}

\newcommand{\addLL}[1]{\textcolor{black}{#1}}

\begin{document}
\title{
Plasmon Resonances and Tachyon Ghost Modes in Highly Conducting Sheets}
\author{D. O. Oriekhov}
\affiliation{Department of Physics, Taras Shevchenko National University of Kiev, Kiev, 03680, Ukraine}
\author{L. S. Levitov}
\affiliation{Massachusetts Institute of Technology, Cambridge, Massachusetts 02139, USA}

\begin{abstract}
  Plasmon-polariton modes in two-dimensional electron gases have a dual field-matter nature that endows them with unusual properties when electrical conductivity exceeds a certain threshold set by the speed of light. In this regime plasmons display an interesting relation with tachyons, the hypothetical faster-than-light particles. While not directly observable,  
tachyons directly impact properties of  
plasmon modes. Namely, in
the ``tachyon'' regime, plasmon resonances remain sharp even when the carrier collision rate $\gamma$ exceeds plasmon resonance frequency. Resonances feature a recurrent behavior as  $\gamma$ increases, first broadening and then narrowing and acquiring asymmetric non-Lorentzian lineshapes with power-law tails extending into the tachyon continuum $\omega>ck$. This unusual behavior 
can be linked to the properties of tachyon poles located beneath $\omega>ck$ 
branch cuts in the complex $\omega$ plane: as $\gamma$ grows, tachyon poles approach the light cone and hybridize with plasmons.
Narrow resonances persisting for $\gamma>\omega$, along with the unusual density and temperature dependence of resonance frequencies, provide clear signatures of the tachyon regime. 
\end{abstract}
\maketitle

\section{Introduction}
\label{sec1}

Surface \addLL{plasmon-polaritons} in atomically thin electron systems feature a number of interesting and potentially useful properties, such as strong light-matter interaction and field confinement, as well as gate tunability\cite{wunsch,hwang,jablan,koppens2011,goncalves2016,basov2016}.  
Plasmon modes, owing to their hybrid charge-field character,  enable powerful \addLL{near-field} diagnostic for electronic properties of two-dimensional (2D) materials\cite{koppens2012,basov2012}. 
The synchronized movement of 
\addLL{charges} in different spatial regions, which constitutes \addLL{plasma oscillations,} is sustained by long-range electron-electron interactions.
In that, the effects of EM retardation due to the finite speed of light are typically small, since electron velocities in solids are nonrelativistic\cite{Giuliani}. 
\addLL{Nonetheless,} since models based on nonretarded Coulomb \addLL{interactions} predict $\omega\sim \sqrt{k}$ dispersion with group velocity diverging at small $k$,
the relativistic retardation effects 
\addLL{inevitably}
become prominent in the long-wavelength limit. Strong retardation  
\addLL{endows the long-wavelength plasmon}
modes 
with novel properties
\addLL{that reflect interesting
dynamical effects inherent to the 3D/2D field-matter binding in this new regime} 
\cite{kukushkin2003,Kukushkin_2015,Muravev_2018}.

Can retardation-dominated 
modes be accessed without changing the plasmon wavelength? This question was first  
posed by Falko and Khmelnitskii\cite{Falko_Khmelnitskii}, who predicted 
enhancement of retardation effects 
upon increasing the conductivity of  
the electron gas. Ref.\onlinecite{Falko_Khmelnitskii} also 
 uncovered a truly puzzling  behavior --- 
collective modes resembling tachyons, the hypothetical superluminal  particles. 
 The regime of interest is reached when 
 the DC ohmic conductivity 
 exceeds the threshold set by the speed of light:
\begin{align}\label{eq:c/2pi}
\sigma >\addLL{\sqrt{\epsilon}}c/2\pi 
,\quad ({\rm in\ SI\ units:}\ \sigma >2\sqrt{\epsilon\epsilon_0/\mu_0})
.
\end{align}
\addLL{with the factor $\sqrt{\epsilon}$ accounting for the dielectric environment (below  we use $\sqrt{\epsilon}=1$ unless stated otherwise).}
In cgs units, used in Eq.\eqref{eq:c/2pi}, ohmic conductivity has dimension of velocity, wherein $2\pi/c\approx 188\,{\rm \Omega}$ per square\cite{resistance_value}. Such values are 
routinely reachable in state-of-the-art 2D electron systems\cite{kukushkin2003,Kukushkin_2015,Muravev_2018}. 
Ref.\onlinecite{Falko_Khmelnitskii}, by analyzing  the dynamics of 2D currents coupled to 3D electromagnetic fields, obtained  modes which, if taken for granted, would describe excitations traveling at superluminal speeds. 
This would of course violate the known laws of physics,
leading to a conclusion that  these are some kind of ghost modes that cannot be observed directly.  Despite several attempts to clarify the meaning of these findings\cite{Volkov_2014,Kukushkin_2015,Muravev_2018}, 
their relation to observable quantities has remained uncertain.

\begin{figure*}
	\centering
\includegraphics[scale=0.27]{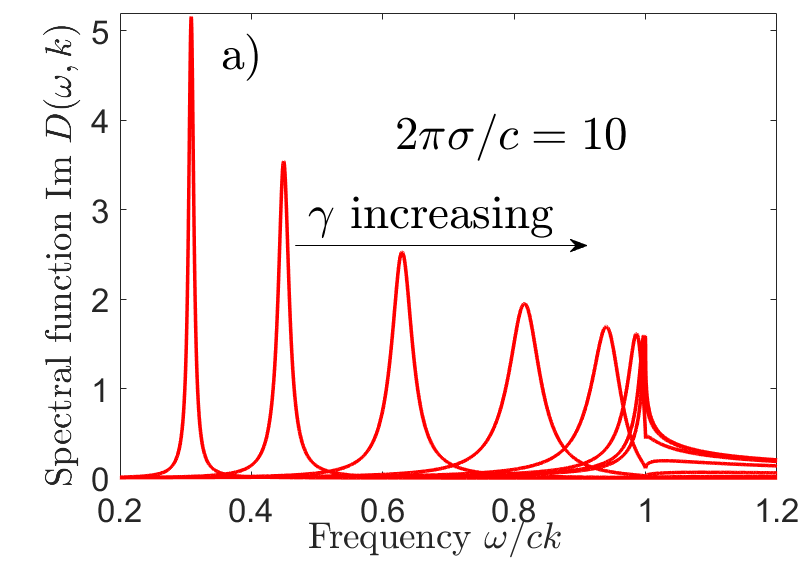}
\includegraphics[scale=0.27]{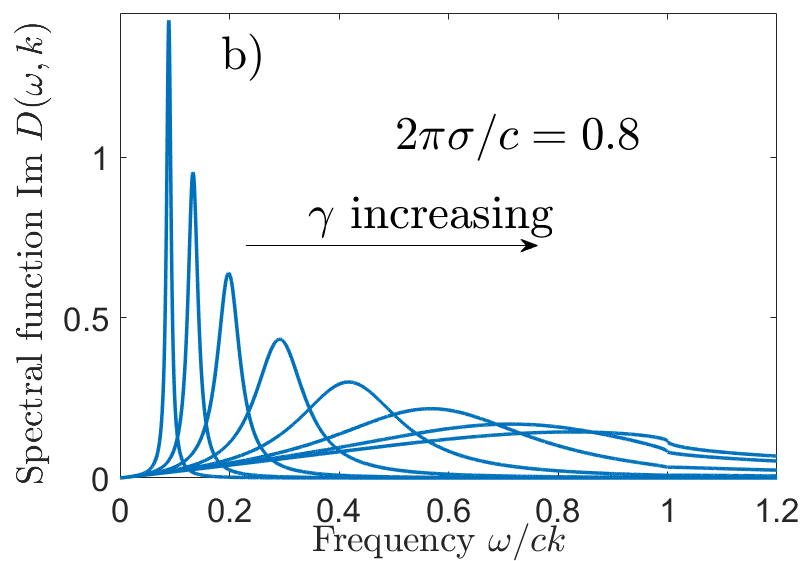}
\includegraphics[scale=0.27]{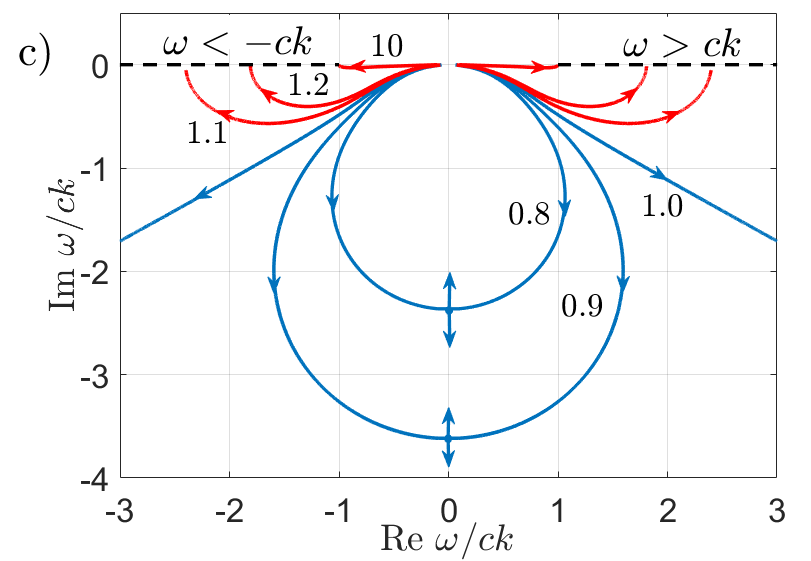}
\caption{a) Recurrent behavior of 
plasmon resonances in the ``tachyon" regime $\sigma >c/2\pi$. Plotted is the dynamic compressibility ${\rm Im}\,D(\omega,k)$, Eq.\eqref{eq:D(w)}, at a fixed $\sigma$. Resonances evolve nonmonotonically as the collision rate $\gamma$ grows, first broadening, then 
sharpening
and developing non-Lorentzian lineshapes, while the resonance frequency  becomes pinned at $\omega\approx ck$ value. b) Non-recurrent behavior at $\sigma<c/2\pi$: resonances broaden and become overdamped as $\gamma$ increases. c) Pole trajectories, 
obtained from Eq.\eqref{eq:poles_Drude} at a fixed $\sigma$.
Arrows show the direction of pole  movement at increasing 
$\gamma$; numbers indicate $2\pi \sigma/c$ values. 
For $2\pi\sigma >c$, 
the poles move under the branch cuts of the square roots in Eq.\eqref{eq:D(w)} 
(dashed lines). Positioned under branch cuts, the poles represent the Falko-Khmelnitskii tachyons, Eq.\eqref{eq:dispersion_superluminal}. The latter, despite being undamped, ${\rm Im}\,\omega=0$, do not generate propagating modes.
}
	\label{fig1}
\end{figure*}

\section{
Plasmon resonances at high collision rates $\gamma\gg\omega$}
\label{sec2}

With this motivation in mind, here we analyze plasmon resonances and their relation to tachyon modes. We focus on  the charge-potential linear response function of a 2D conducting sheet, $\rho_{\omega,k}=-D(\omega,k) \phi_{\omega,k}$.
The dynamical compressibility $D(\omega,k)$
is found to be expressed through the dispersive sheet conductivity $\sigma(\omega)$ as
\be\label{eq:D(w)}
D(\omega,k)=\frac{k^2\sigma(\omega)}{2\pi q(\omega)\sigma(\omega)-i\omega}
,\quad
q(\omega)=\sqrt{k^2-\frac{\omega^2}{c^2}}
.
\ee
	The conductivity, in general, depends also on the wave number $k$ and the ee scattering rates. However, these effects are important only at relatively large values $k\sim \omega /v_F$,  
	whereas the values  relevant in our case are much smaller: $k\sim \omega / c$. 
The dielectric constant of the surrounding medium, ignored here for simplicity,
will be accounted for below, see Eq.\eqref{eq:D(w)_epsilon}.

The spectral function ${\rm Im}\,D(\omega,k)$ 
describes plasmon resonances in several different regimes. At $\sigma >c/2\pi$, the resonances acquire an interesting recurrent character, which is illustrated in Fig.\ref{fig1}. As the collision rate $\gamma$ grows, with the conductivity $\sigma$ and wavenumber $k$ values kept fixed, resonances first broaden, but then, when $\gamma$ exceeds $ck$, they begin to 
sharpen as $\gamma$ increases. Simultaneously, resonance frequency becomes pinned at $\omega=ck$ value and lineshapes change from Lorentzian to highly non-Lorentzian. Strikingly, resonances remain sharp even when the collision rate $\gamma$ is much greater than the resonance frequency $\omega$. In this regime, lineshapes become asymmetrical, cuspy, and develop tails extending far in the $\omega>ck$ continuum. At $\sigma <c/2\pi$, on the contrary, a conventional behavior takes place: resonances broaden and weaken as $\gamma$ grows. 

The physical reason for resonances 
sharpening can be understood as a  reduction in damping due to a change in the mode makeup upon frequency approaching $ck$. Indeed, at $\omega<ck$ the field outside the conducting sheet represents an evanescent wave decaying as a function of distance  as $e^{-\lambda z}$ with the decay parameter $\lambda = q(\omega)$. Since the latter becomes small as $\omega$ approaches $ck$, the mode confinement in the direction perpendicular to the plane becomes less tight, leading to an enhancement in the field-matter volume ratio. This makes the mode overlap with two-dimensional electrons smaller and, therefore, reduces damping. Here and below we assume that dissipation is dominated by ohmic losses of 2D electrons; the situation in experimental systems can be more complicated due to losses in the surrounding medium.

Plasmon resonances that sharpen 
\addLL{even though} 
the collision rate $\gamma$ exceeds resonance frequency also suggest an 
interpretation in terms of motional narrowing. 
A resonant frequency that has a smaller linewidth than may be expected, is a common behavior in systems where oscillations occur in the presence of a rapidly changing environment. The motional narrowing effect arises due to the changes quickly averaging out in accordance with the central limit theorem, and therefore decoupling 
from the oscillating degrees of freedom. For plasmon resonances, motional narrowing is often regarded as a signature of the hydrodynamic regime in which plasmon excitation is shared among many particles that quickly exchange their microscopic states through two-body scattering. 
In contrast, the present problem 
\addLL{features} 
motional narrowing \addLL{that} results
from \addLL{oscillations 
supported} by a large number of quickly relaxing degrees of freedom, producing resonances that remain sharp even at high collision rates $\gamma\gg\omega$. 

\begin{figure*}[t]
	\centering
	\includegraphics[scale=0.35]{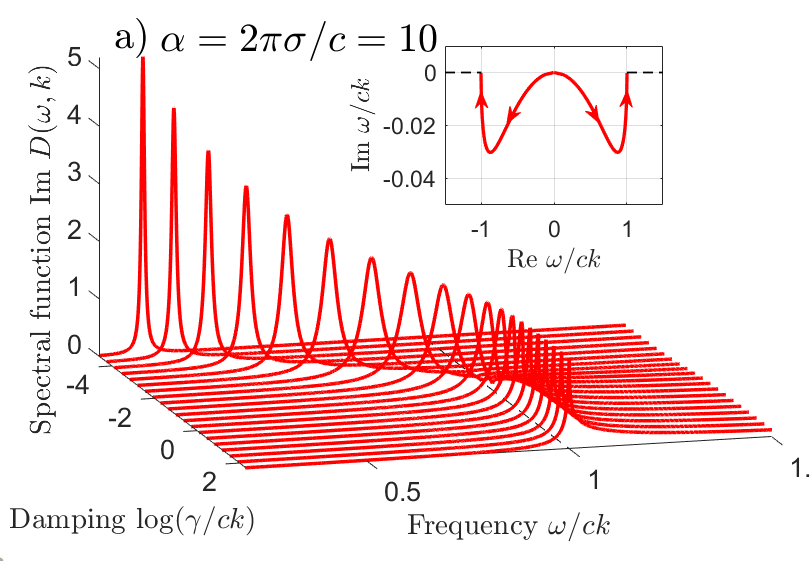}
	\includegraphics[scale=0.35]{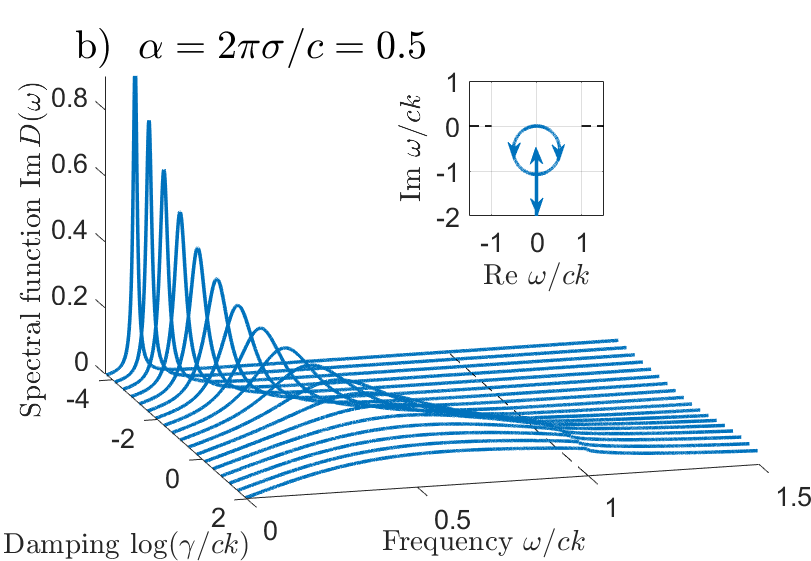}
	\caption{Frame-by-frame 
		evolution of the resonances in Fig.\ref{fig1}. a) 
		At a constant $\sigma>c/2\pi$, resonances 
		are dispersing as the collision rate grows 
from $\gamma\ll ck$ to $\gamma\sim ck$; \addLL{the dispersion is quenched for $\gamma>ck$. In the latter case the non-dispersing}
resonance frequency becomes pinned at the edge of the continuum $\omega>ck$, taking \addLL{values} $\omega\approx ck$. 
\addLL{The dependence of resonance width vs. $\gamma$}
is nonmonotonic, broadening while $\gamma<ck$ and narrowing once $\gamma$ exceeds $ck$. 
		b) Conventional behavior at  $\sigma<c/2\pi$: resonances broaden as $\gamma$ grows and are washed out once $\gamma$ exceeds the resonance frequency. 
		Insets in a) and b) show the trajectories of $D(\omega,k)$ poles in the complex $\omega$ plane, obtained by varying $\gamma$. 
	}
	\label{fig2}
\end{figure*}

In experiment, the key system parameters -- $\gamma$ and $\sigma$ -- can be varied independently by tuning temperature and carrier density ($T$ and $n$). However, since in general the $T$ and $n$ dependence of $\gamma$ and $\sigma$ may be fairly complicated, here it will be convenient to view these quantities as
proxies for the experimental knobs, treating them as independently tunable variables. This represents a meaningful choice also because the quantity $\sigma$ is directly measurable, and thus the recurrent evolution of resonances at a fixed $\sigma$ and varying $\gamma$, such as that shown in Figs.\ref{fig1} and \ref{fig2}, can be extracted directly from the measurement results without knowing the exact dependence of $\gamma$ and $\sigma$ on the experimental knobs such as $T$ and $n$.

Quantitative estimates suggest 
that the regime of interest is readily accessible in atomically-thin  materials currently under investigation in 
nanoscale plasmonics \cite{wunsch,hwang,jablan,koppens2011,goncalves2016,basov2016,koppens2012,basov2012}. Namely, in graphene, 
the carrier mean free path can be as large as $10$-$20\,{\rm \mu m}$, exceeding by a large margin the values $\sim 1\,{\rm \mu m}$ set by the threshold in Eq.\eqref{eq:c/2pi}. 
These aspects 
are discussed in greater detail in Secs.\ref{subsec-tachyon:numerical},\ref{sec3}.

It is also interesting to mention that superluminal modes somewhat reminiscent of our tachyons have appeared previously in the literature on the surface Zenneck wave problem. 
The Zenneck wave propagates at the surface of a lossy medium in a three-dimensional space (see Refs.\onlinecite{barlow1953, michalski2015, babicheva,oruganti2020} and references therein); Maxwell equations describing these waves admit solutions with superluminal dispersion $\omega=c'k$, $c'>c$. However, the puzzling superluminal aspects aside, our ghost modes are distinct from those in the Zenneck problem. One difference is dimensionality (2D vs. 3D), another is the character of the EM field -- residing within the lossy medium vs. the free space outside the conducting sheet, respectively. Even more important is the different character of the observable. The Zenneck wave 
can manifest itself 
through resonances with radiation incident from 3D at a certain angle \cite{barlow1953,babicheva}. In contrast, our modes represent
resonances in the dynamical response functions. The relation between our problem and the Zenneck wave problem will be discussed elsewhere. 

\section{Resonance sharpening due to tachyon poles near the light cone}
A unique insight into the properties of the resonances, in particular their relation with the tachyon modes of Ref.\onlinecite{Falko_Khmelnitskii}, can be gained by investigating 
complex-$\omega$ poles of $D(\omega,k)$. Here we focus on 
the dispersive Drude model: 
\begin{align}\label{eq:dispersion_poles}
i\omega=2\pi q(\omega) \sigma(\omega)
,\quad
\sigma(\omega)=\frac{ne^2}{m(\gamma-i\omega)}
,
\end{align} 
The dispersion relation \eqref{eq:dispersion_poles}, after simple algebra, yields a characteristic equation
\begin{align}\label{eq:poles_Drude}
\frac{\omega^2(\omega+i\gamma)^2}{\beta^2} +\frac{\omega^2}{c^2}=k^2
,\quad
\beta=\frac{2\pi ne^2}{m}
.
\end{align}
The complex roots of this quartic equation can be found explicitly.
Two of these roots are the poles of $D(\omega,k)$ 
shown in Fig.\ref{fig1}(c). Two spurious roots, added when the square root in $q(\omega)$ is rationalized, are discarded.

The behavior of poles in the complex $\omega$ plane, which is illustrated in Fig.\ref{fig1}(c), mimics the recurrent behavior of resonances in the ``tachyon" regime $\sigma>c/2\pi$:  as $\gamma$ increases and $\sigma$ is kept fixed, the poles first move away from the real $\omega$ axis, then make a U-turn and  move back towards the real axis, landing on the lower side of the branch cuts $\omega<-ck$ and $\omega>ck$. Likewise, at $\sigma<c/2\pi$ pole trajectories show a non-recurrent behavior: 
moving gradually away from the real axis without turning back, and then colliding at the imaginary axis to create a pair of overdamped modes with pure imaginary $\omega$. 

Quantitatively, this behavior can be described most easily by taking the limit $2\pi\sigma\gg c$ in Eq.\eqref{eq:poles_Drude}. In this case, as Fig.\ref{fig1} suggests, the real part of $\omega$ is much greater than the imaginary part. Ignoring the latter at first, we take the $\omega\gg\gamma$ limit. This yields a dependence 
\be
k^2=\frac{\omega^4}{\beta^2}+\frac{\omega^2}{c^2}
,
\ee
which gives the dispersion $\omega=(\beta k)^{1/2}$ at large $k$, and a light-like dispersion $\omega=ck$ at small $k$, as expected. The imaginary part of $\omega$, which provides an estimate for resonance width,  can be found by replacing $\omega\to\omega-i\Gamma$, and expanding in small $\gamma$ and $\Gamma$ to first order. This gives
\be\label{eq:Gamma_gamma}
\Gamma=\frac{\omega^2\gamma}{2\omega^2+\frac{\beta^2}{c^2}}
.
\ee
Substituting $\beta=2\pi\sigma\gamma$ and taking $\sigma$ to be constant, we see that Eq.\eqref{eq:Gamma_gamma} predicts a nonmonotonic dependence for $\Gamma$ vs. $\gamma$. For the resonance width, estimated as $\Gamma$, this behavior is in good agreement with the recurrent evolution of resonances and poles at varying $\gamma$ and constant $\sigma$, as shown in Fig.\ref{fig1} and, in greater detail, in Fig.\ref{fig2}. 

\begin{figure*}
	\centering
	\includegraphics[scale=0.37]{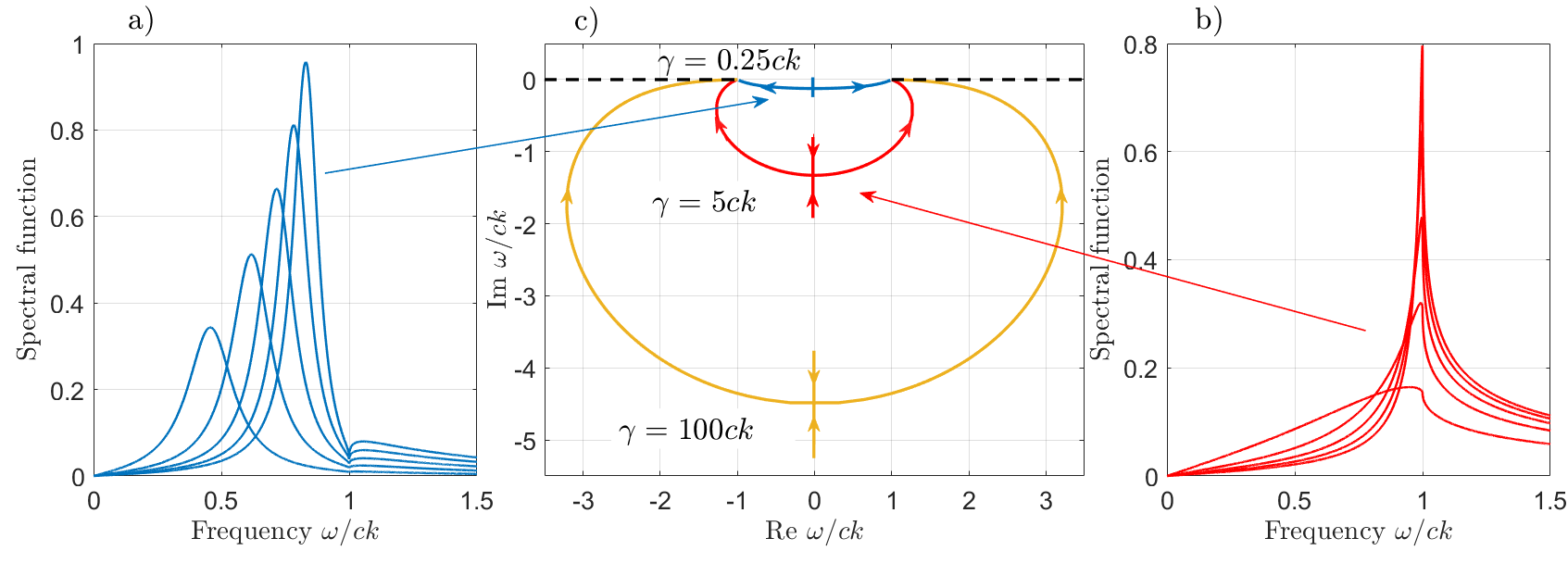}
	\caption{a),b) Evolution of resonances at a fixed $\gamma$ and $2\pi\sigma/c$ increasing from $1$ to $5$. Resonances sharpen as $\sigma$ increases; in b) they 
\addLL{sharpen and narrow down} despite that the resonance frequency remains smaller than $\gamma$. c) Pole trajectories found from  
		Eq.\eqref{eq:poles_Drude} plotted at a fixed $\gamma$ for 
		$2\pi\sigma/c$ varying 
		from $0.01$ to $20$. The direction of pole movement is indicated by arrows.
		The branch cuts 
		$|\omega|>c|k|$ are shown by dashed lines; $\gamma$ values are indicated next to arrows. 
		Tachyon resonances arise at large $\sigma$ as the poles approach the branch cuts. 
	}
	\label{fig:poles_gamma}
\end{figure*}

At $2\pi\sigma\gtrsim c$ and high damping $\gamma$, the poles of $D(\omega,k)$ are positioned directly beneath the branch cuts (see Fig.\ref{fig:poles_gamma}). In the limit $\gamma\gg\omega$, after approximating $\sigma(\omega)\approx \sigma+\frac{i\omega}{\gamma}\sigma$, simple algebra gives 
\be\label{eq:dispersion_superluminal}
\omega_{\pm}=\pm vk-i\gamma'
,\quad
v
=\frac{c\alpha}{\sqrt{\alpha^2-1}}
,\quad 
\alpha=\frac{2\pi\sigma}{c}
,
\ee
the values identical to those found in Ref.\onlinecite{Falko_Khmelnitskii}, 
with damping $\gamma'=\frac{\alpha^2 c^2 k^2}{\gamma (\alpha^2-1)^2}$ vanishing at high $\gamma$. As noted in Ref.\onlinecite{Falko_Khmelnitskii}, the peculiar dispersion relation with greater-than-$c$ group velocity does not imply superluminal signal propagation. The reasons for that, which are somewhat subtle, can be 
summarized as follows.

First, since at large $\gamma$ the frequencies $\omega_{\pm}$ reside directly at the branch cuts $\omega>ck$ and $\omega<-ck$, the poles $\omega=\omega_{\pm}$ do not represent isolated singularities; rather the poles and branch cuts must be handled jointly as {\it compound}, or {\it unseparable,} singularities. Another point of note, which is more essential than the ``compound singularity'' property, is that the poles reside on the lower (unphysical) sides of the branch cuts, which separate the poles from the upper imaginary halfplane ${\rm Im}\,\omega>0$. Since it is the $\omega$ dependence in that halfplane that governs time evolution of a response, the poles separated from the ${\rm Im}\,\omega>0$ domain by branch cuts cannot create, on their own, any $v>c$ modes. More formally, below we demonstrate that these poles give no singular contributions to the spectral function because their residues vanish, see Eq.\eqref{eq:D(w)_residue} and accompanying discussion. Instead, the poles under the branch cuts alter the shapes of the resonances positioned at $\omega\lesssim ck$, which remain sharp even when $\gamma\gg\omega$ but acquire asymmetric line shapes with the tails 
extending into the tachyon continuum $\omega>ck$.

\section{Experimental accessibility of the ``superluminal'' regime $\sigma > c/2\pi$} 
\label{subsec-tachyon:numerical}
Here we provide quantitative estimates which illustrate that the regime $\sigma > c/2\pi$ is readily accessible in atomically thin conductors such as graphene monolayer and bilayer. To facilitate estimates, we write Drude conductivity as
\[
\sigma=\frac{ne^2}{m}\tau=\frac{g_v g_s}{4\pi}\frac{e^2}{\hbar}k_F\ell
\]
where $k_F$ and $\ell$ are the electron Fermi momentum and mean free path values, and $g_s=g_v=2$ describes the spin and valley degeneracy. Using this result, the relation $\sigma > c/2\pi$ can be written as a condition for the mean free path:
\be\label{eq:ell>c}
\ell>\frac{\hbar c}{e^2}\frac1{2k_F}
.
\ee
Evaluating $k_F$ for a typical carrier concentration $n=10^{12}\,{\rm cm^{-2}}$ we obtain the value $\frac1{2k_F}\approx 9\,{\rm nm}$. Multiplying this result by $\frac{\hbar c}{e^2}\approx 137$ brings the condition in Eq.\eqref{eq:ell>c} to the form
\[
\ell> 1.2\,{\rm \mu m}
.
\]
However, the mean free path values in high-mobility graphene monolayer and bilayer  routinely reach $10$-$20\,{\rm \mu m}$, which is comfortably in the range set by the bound in Eq.\eqref{eq:ell>c}. 
This indicates that the condition $\sigma>c/2\pi$ can be easily met. 

The condition $\sigma>c/2\pi$ can also be achieved in 
metallic few-atom-thin films, a system where surface plasmons have been investigated recently\cite{Fattah2019}. In thin films, the carrier mean free path is limited by the film thickness, i.e. is relatively short.  However the carrier density in films is much greater than in graphene. For example, for a few-nanometer-thin film the effective 2D carrier density is on the order $n\approx 10^{17}\,{\rm cm^{-2}}$ whereas $\ell\approx 1\,{\rm nm}$. Comparing to the above we see that the shorter mean free path value is balanced by the larger $k_F$ value, so that the condition $\sigma>c/2\pi$ remains reachable.

\section{The ``superluminal'' plasmonic response} 
\label{sec3}

To validate the picture discussed in Secs.\ref{sec1} and \ref{sec2}, we consider the charge-potential response 
in the time domain: 
\be
\rho_k(t)=-\int_{-\infty}^\infty dt' D_k(t-t')\phi_k(t')
,
\ee
corresponding to $\rho_{\omega,k}=-D(\omega,k)\phi_{\omega,k}$ 
at a fixed wavenumber $k$. 
The memory function
$D_k(t-t')$ equals
\be\label{eq:memory_fnc}
D_k(\tau)=\int_{-\infty}^\infty \frac{d\omega}{2\pi} e^{-i\omega\tau}D(\omega,k)
.
\ee
Here the integral runs over a straight path $-\infty<\omega<\infty$ just above the real axis. The causality condition $D_k(\tau<0)=0$ is ensured, as always, by analyticity of $D(\omega,k)$ in the upper halfplane ${\rm Im}\,\omega>0$. 

To see why the expression in Eq.\eqref{eq:D(w)}, when plugged in  Eq.\eqref{eq:memory_fnc}, does not generate propagating modes with $v>c$,  
we start with a simple technical observation regarding analytic properties of $q(\omega)$. The quantity $q(\omega)$ is real in the domain  $-ck<\omega<ck$ and pure imaginary at $\omega>ck$ and $\omega<-ck$ with a sign that must be determined by analytic continuation. The recipe for 
continuation follows from analyticity of $ q(\omega) $ 
in the halfplane ${\rm Im}\,\omega>0$, prescribed by causality. 
Therefore, $q(\omega)$ should be treated as $\sqrt{k^2-\frac{(\omega+i0)^2}{c^2}}$ with an infinitesimal positive shift in $\omega$, giving 
\begin{equation}
\label{kappa_analytic_cont}
q(\omega)=\begin{cases}
\sqrt{k^2-\frac{\omega^2}{c^2}},& -ck<\omega<ck\\
-i\,{\rm sgn}\,\omega\sqrt{\frac{\omega^2}{c^2}-k^2},& \omega<-ck,\  \omega>ck
, 
\end{cases}
\end{equation} 
where the sign factor $-{\rm sgn}\,\omega$ for  the cases 
$\omega>ck$ and $\omega<-ck$ arises due to analytic continuation through the upper halfplane.  A simple consequence of this result is that the dispersion equation obtained in Ref.\onlinecite{Falko_Khmelnitskii} does not have solutions 
at the real  axis on the 
upper side of branch cuts. The solutions given in \eqref{eq:dispersion_superluminal} are located under the cuts $\omega>ck$ and $\omega<-ck$. Therefore, from the point of view of analytical properties, they 
represent fictitious poles, or more precisely, the poles located on a non-physical sheet of the Riemann surface of complex frequency $\omega$. As such, they do not generate propagating modes.

This point can be illustrated by transforming the expression in Eq.\eqref{eq:D(w)} in the $\gamma\gg\omega$ limit to the form
\be\label{eq:D(w)_residue}
D(\omega,k)=
\frac{\alpha ck^2(\alpha \sqrt{c^2k^2-\omega^2}+i\omega)}{
\alpha^2 c^2 k^2-(\alpha^2-1)\omega^2}
,
\ee
where we replaced $\sigma(\omega)$ in Eq.\eqref{eq:D(w)} by $\sigma=c\alpha/2\pi$, and rationalized denominator by multiplying it by $\alpha \sqrt{c^2k^2-\omega^2}+i\omega$. This expression has poles on the real axis at the tachyon frequencies $\omega=\pm vk$ with $v>c$, Eq.\eqref{eq:dispersion_superluminal}, so long as $\alpha>1$. However these poles give a vanishing contribution to the spectral function evaluated at ${\rm Im}\,\omega=+i0$ because the numerator, owing to the sign prescription found above, Eq.\eqref{kappa_analytic_cont},  vanishes at the poles. As a result, the spectral function is smooth at the tachyon frequencies $|\omega|>ck$. This is clearly seen in the resonances shown in Figs.\ref{fig1} and \ref{fig2} which have smooth tails extending into the tachyon continuum with cusps at $\omega=ck$ but no singularities at $\omega>ck$. 

Next, we proceed to derive the response function given in Eq.\eqref{eq:D(w)} and estimate the relevant experimental parameter values.  
We start with EM equations in 3D space due to 2D currents, for generality adding a dielectric constant of the surrounding medium. Using Fourier harmonics, in Lorentz gauge we have $\vec k\cdot\vec A_{\vec k,\omega}-\frac{\omega}{c}\epsilon \phi_{\vec k,\omega}=0$, and 
\begin{align}\nonumber
& (\vec k^2-{\textstyle \frac{\omega^2}{c^2}}\epsilon)\vec A_{\vec k,\omega}=\frac{4\pi}{c}\vec j_{\vec k,\omega}
,\quad  (\vec k^2-{\textstyle \frac{\omega^2}{c^2}\epsilon})\phi_{\vec k,\omega}=\frac{4\pi}{\epsilon}\rho_{\vec k,\omega}
.
 \end{align}
Taking $z$ axis to be perpendicular to the 2D sheet, and working in a mixed Fourier representation, 
\[
\phi_{\vec k,\omega}(z)=\sum_{k_z}\phi_{\vec k,\omega}e^{ik_z z}
,\quad
\rho_{\vec k,\omega}(z)=\sum_{k_z}\rho_{\vec k,\omega}e^{ik_z z}
,
\] 
where from now on $\vec k$ is two-dimensional, we have
\begin{align}\label{eq:phi_rho}
&(\partial^{2}_{z}-k^2+{\textstyle \frac{\omega^2}{c^2}}\epsilon)\phi_{\vec k,\omega}(z)=-\frac{4\pi}{\epsilon}\rho_{\vec k,\omega}\delta(z),\\
\label{eq:A_j}
&(\partial^{2}_{z}-k^2+{\textstyle \frac{\omega^2}{c^2}}\epsilon)\vec{A}_{\vec k,\omega}(z)=-\frac{4\pi}{c}\vec{j}_{\vec k,\omega}\delta(z).
\end{align}
Solving Eqs.\eqref{eq:phi_rho} and \eqref{eq:A_j} for the $z$ dependence gives
\be
\label{eq:potentials}
\phi_{\vec k,\omega}(z)=\frac{2\pi \rho_{\vec k,\omega}}{\epsilon q(\omega)}e^{-q(\omega)|z|},\quad 
\vec{A}_{\vec k,\omega}(z)=\frac{2\pi\vec{j}_{\vec k,\omega}}{cq(\omega)}e^{-q(\omega)|z|}
, 
\ee
with $q(\omega)=\sqrt{k^2-\frac{\omega^2}{c^2}\epsilon}$.

These relations must be combined with the conductivity response 
$
{\vec j}'=\sigma(\omega)\vec E
$.
Here
the prime indicates the induced current, whereas the quantities in Eq.\eqref{eq:potentials} should be taken as sums of the external and induced contributions, $\rho=\rho'+\rho_{\rm ext}$, $\vec j=\vec j'+\vec j_{\rm ext}$. Writing
$\vec{E}=-\nabla\phi-\frac1{c}\partial_{t}\vec{A}$ and using the continuity relations for the 2D currents and charges, $\rho_{k,\omega}=\frac{1}{\omega}\vec{k}\cdot\vec{j}_{k,\omega}$, 
we eliminate variables $\rho$ and $\phi$ to obtain
\begin{align}
\vec j'_{\vec k,\omega}=i\omega\frac{2\pi\sigma(\omega)}{q(\omega) c^2}  \lp\vec j_{\vec k,\omega} 
-\frac{c^2}{\omega^2\epsilon}\vec k\lp \vec k\cdot\vec j_{\vec k,\omega} 
\rp\rp
,
\end{align}
with $\vec j=\vec j'+\vec j_{\rm ext}$. This relation can be put in the form of a $2\times 2$ matrix response function, $\vec{j}'_{\vec{k},\omega}=M\lp \vec{j}'_{\vec k,\omega}+\vec{j}^{\rm ext}_{\vec k,\omega}\rp $. For longitudinal waves $\vec{j}_{\vec k,\omega}\parallel\vec{k}$, $\vec{j}_{\vec k,\omega}^{\rm ext}\parallel\vec{k}$ we obtain 
\be
\vec{j}_{\vec{k},\omega}=\frac1{1-M}\vec{j}^{\rm ext}_{\vec k,\omega}
=\frac{i \omega\epsilon }{i \omega\epsilon -2 \pi  q(\omega)  \sigma(\omega) }
\vec{j}^{\rm ext}_{\vec k,\omega}.
\ee 
Dynamical compressibility can now be 
found by substituting in place of $\vec{j}^{\rm ext}_{\vec k,\omega}$ the current induced by an external potential, $-i\sigma(\omega)\vec k \phi^{\rm ext}_{k,\omega}$. Relating the net current $\vec j=\vec j'+\vec j_{\rm ext}$ to the net charge as $\rho=\frac1{\omega}\vec k\cdot\vec j$ gives 
\be\label{eq:D(w)_epsilon}
\rho_{\omega,k}=\frac{k^2\sigma(\omega)\epsilon}{i\omega\epsilon-2\pi q(\omega)\sigma(\omega)}\phi^{\rm ext}_{\omega,k}
,
\ee
which is the result in Eq.\eqref{eq:D(w)} generalized to $\epsilon\ne 1$. 
As a sanity check, at $\omega=0$ we recover the standard result for an ideal conductor $\rho_{k}=-\frac{\epsilon k}{2\pi}\phi_{k}$, where the minus sign describes perfect screening of an external potential by induced charges. 

The result in Eq.\eqref{eq:D(w)_epsilon} can be related to the $\epsilon=1$ result in Eq.\eqref{eq:D(w)} by absorbing $\epsilon$ into rescaled parameters, 
\be
c\to \tilde c=\frac{c}{\sqrt{\epsilon}}
,\quad
\sigma\to \tilde\sigma=\frac{\sigma}{\epsilon}
,
\ee
upon which the dimensionless ratio $\alpha=2\pi\sigma/c$ is reduced by a factor $\sqrt{\epsilon}$. Accounting for this change, the results above can be applied directly, with the condition in Eq.\eqref{eq:c/2pi} replaced by $\sigma>\frac{\sqrt{\epsilon}c}{2\pi}$, and so on.
For a system of size $L=20\,{\rm \mu m}$, using the value $\epsilon\approx 11$ (sapphire), the resonance frequency is $\omega_0=\frac{\pi c}{\sqrt{\epsilon}L}=2\pi\times 2.26\,{\rm THz}$. This value can be reduced by using proximal gates to screen the electron-electron interactions.

\section{Discussion and conclusions}


\addLL{Sharp plasmon resonances, occurring despite} 
the collision rate exceeds the resonance frequency, $\gamma\gg\omega$, is a striking behavior that can be attributed to motional  narrowing due to \addLL{the}
many quickly relaxing microscopic degrees of freedom that plasmon excitations are made of, \addLL{which is set by the carrier density $n$. As demonstrated above, an increase in $n$ overwhelms an increase in $\gamma$, producing sharper resonances when conductivity, which is proportional to $n/\gamma$, exceeds the $c/2\pi$ threshold.} Motional narrowing of collective modes is of course familiar in the hydrodynamic regime, taking place in plasmonics when plasmon frequency is smaller than the electron-electron scattering rate, $\omega\ll\gamma_{\rm ee}$. Here we encounter a more exotic behavior: resonance sharpening through motional narrowing arising due to rapid momentum-relaxing collisions. It is usually taken for granted that high collision rates $\gamma\gg\omega$ produce rapid damping that broadens plasmon resonances. However, as the discussion above 
\addLL{shows,} this simple intuition fails for electron systems with conductivity taking high ``superluminal'' values \addLL{$\sigma>c/2\pi$.} In this case, perhaps somewhat counterintuitively, rapid relaxation 
\addLL{does not present 
an obstacle to the formation of}  
abnormally narrow resonances. 

This surprising behavior can also be linked to the peculiar evolution of \addLL{the} poles of the response function in the complex frequency plane. At small $\gamma$ the poles represent the conventional 
collisionless plasmons. As $\gamma$ grows, the poles move under the branch cuts, turning into tachyon modes with faster-than-$c$ group velocity, first predicted by Falko and Khmelnitskii. Since the poles are positioned on the unphysical sheet of the complex frequency Riemann surface, they do not result, by themselves, 
in propagating modes. However, as these superluminal poles approach the light cone $\omega=ck$, they influence the observable response by producing plasmon resonances with distinct non-Lorentzian lineshapes and sharpening them despite the collision rate being high. These features, along with a characteristic nonmonotonic dependence on experimental knobs, 
provide clear signatures of the tachyon regime. 
The relation between tachyon poles and 
plasmonic resonances that sharpen when conductivity increases above the threshold value set by the speed of light can therefore be useful as a way to probe the elusive tachyon modes. 


We are grateful to I. V. Kukushkin and K. E. Nagaev for useful discussions. L.L. acknowledges support from the Science and Technology Center for Integrated Quantum Materials, NSF Grant
No. DMR-1231319. Part of this work was performed at the Aspen Center for Physics, which is supported by National Science Foundation grant PHY-1607611.


\begin{thebibliography}{99}
\bibitem{resistance_value} Resistance value in Eq.\eqref{eq:c/2pi} can be estimated by combining the fine structure constant $\alpha=e^2/\hbar c\approx 1/137$ and the von Klitzing resistance quantum $h/e^2=25.8\,{\rm k\Omega}$. The quantity $2\pi/c$ then translates into resistance of $2\pi (e^2/\hbar c)(\hbar/e^2)\approx 188\,{\rm \Omega}$ per square.
\end{thebibliography}
\end{document}